# Highly-cited papers in Library and Information Science (LIS): Authors, institutions, and network structures

*Journal of the Association for Information Science and Technology* (in press)

Johann Bauer[1]*, Loet Leydesdorff[2] and Lutz Bornmann [3]

[1] Max Planck Institute of Biochemistry, Am Klopferspitz 18, 82152 Martinsried, (Germany); jbauer@biochem.mpg.de; *corresponding author.

[2] Amsterdam School of Communication Research (ASCoR), University of Amsterdam, P.O. Box 15793, 1001 NG Amsterdam (The Netherlands); loet@leydesdorff.net

[3] Division for Science and Innovation Studies, Administrative Headquarters of the Max Planck Society, Hofgartenstr. 8, 80539 Munich (Germany); bornmann@gv.mpg.de

**Abstract**

As a follow-up to the highly-cited authors list published by Thomson Reuters in June 2014, we analyze the top-1% most frequently cited papers published between 2002 and 2012 included in the Web of Science (WoS) subject category "Information Science & Library Science." 798 authors contributed to 305 top-1% publications; these authors were employed at 275 institutions. The authors at Harvard University contributed the largest number of papers, when the addresses are whole-number counted. However, Leiden University leads the ranking, if fractional counting is used.

Twenty-three of the 798 authors were also listed as most highly-cited authors by Thomson Reuters in June 2014 (http://highlycited.com/). Twelve of these 23 authors were involved in publishing four or more of the 305 papers under study. Analysis of co-authorship relations among the 798 highly-cited scientists shows that co-authorships are based on common interests in a specific topic. Three topics were important between 2002 and 2012: (1) collection and exploitation of information in clinical practices, (2) the use of internet in public communication and commerce, and (3) scientometrics.

## 1. Introduction

Citation counts can be considered an indicator of the impact a paper has. Papers attracting a large number of citations are of interest, because their citation suggests a particular influence on the development of science. However, using citation counts as a base of research evaluation requires normalization, because differences in time intervals for attracting citations, different citation patterns in research fields, and different coverage of the discipline-specific literature in multi-disciplinary databases bias the numbers (Levitt & Thelwall, 2009; Podlubny & Kassayova, 2006). Percentiles of citations have been proposed for normalizing citations (Bornmann, Leydesdorff, & Mutz, 2013). They are especially appropriate to identify excellent papers, because they can be used to indicate the ranks of papers in their database years and in their categories as defined in the Essential Science Indicators (ESI) or the Web of Science (WoS) (Bornmann, 2014). Papers considered to be excellent allow the analysis of the affiliations of highly cited authors as well as topics prevailing in journals of a WoS category of interest at the time of analysis (Bornmann, 2014; Abramo, et al., 2014; Tang 2004).

The more science policy focuses on research excellence, the more researchers and institutions are confronted with evaluations. In June 2014, Thomson Reuters published its new (2014) list of 3,215 highly-cited researchers at http://highlycited.com/. The highly-cited list is based on the number of the top-1% most highly-cited papers per author in the (eleven) years 2002-2012. The percentile ranks were normalized using the 22 so-called broad categories for journals in the ESI as reference sets. The ESI-database is virtually similar to the WoS in journal composition; but journals in WoS are categorized using the 251 so-called WoS-categories among which one is for "Information Science & Library Science" (LIS). This set refers to 83 journals.

In this study we selected the 1% most highly-cited papers of the LIS category (2002-2012). Then we determined authors and institutions contributing to the papers with the aim to further investigate the top-layers of authors and papers in terms of percentile ranks for the LIS category in WoS. In order to distinguish specificities of citation and co-authorship patterns among elite scientists compared to more typical scientists (Bornmann, de Moya Anegón, & Leydesdorff, 2010), we compared the list of highly-cited authors and their affiliations based on the LIS category in WoS with the ESI based list published at http://highlycited.com/. Furthermore, we generated networks among both the authors and the institutions involved. We found that trends in information research are set by authors (producing highly-cited papers) of collaboration networks.

## 2. Methods

*2.1. Data set used*

The analytical version of WoS installed at the Max Planck Digital Library (MPDL) in Munich provides normalized percentile ranks per WoS category, publication year, and document type. The percentile ranks are calculated using the method of Hazen (1914, p. 1550). Querying this database for the period of 2002-2012 provided us with 30,450 documents in LIS (871 reviews and 29,579 articles) in June, 2014.[1] Of these papers, 305 were in the top-1% percentile rank.

*2.2. Disambiguation of author names*

Authors were identified by names, initials and affiliations. Some authors used only their first initial in some papers, but in other papers more than a single initial were used. If the first name of an author, the surname, and the affiliation were equal, these author names were considered to identify a single person i.e. a unique author.

*2.3. Unifications of institutions*

All institutions provided by the authors of the papers were identified. Details about departments, working groups, etc., were not considered. Furthermore we unified variants of institution names (e.g. Harvard University and Harvard Medical School) and combined all individual institutions of an organization—insofar as they could be recognized. Thus, for example, we combined all individual universities of the University of California System as well as the various subunits of the NIH as in ESI. A number of the highly-cited researchers are affiliated not just to a single, but up to three different institutions. For this reason, we generated institutional ranking lists, which are based either on only the first-named institution by each author or on all the institutions (primary and further addresses) named by the authors (Bornmann & Bauer, 2014).

*2.4. Generation of networks*

Networks were generated on the basis of co-authorship relations among the 798 disambiguated author names and at the aggregated level among the 275 unified institutional addresses using txt2Paj.exe. txt2Paj.exe at http://www.pfeffer.at/txt2pajek can be used to transform an Excel file in csv-format (comma separated variables) into the .net-format of Pajek[2] (Leydesdorff, Khan, & Bornmann, 2014). For the purpose of visualization, the resulting Pajek files were separated into eight groups of ten or more co-authors (n = 347 of the total of 798 authors) and two groups of four or more institutions (n = 186 of the total of 275 institutions) co-authoring within the set. The algorithm of Kamada & Kawai (1989) is used for the layout of the networks.

## 3. Results

*3.1. Authors contributing to the 305 top-1% highly-cited papers*

After disambiguation of author names we counted 798 individuals, who authored or co-authored the 305 papers under study. Among these 798 authors of top-1% papers in the LIS category of WoS we found a match with 23 authors (2.9%) listed among the most highly-cited authors at http://highlycited.com/ (Table 1). These authors are assigned by Thomson Reuters to the ESI categories "Social Science, general" (17), "Economics & Business" (3), "Computer Science" (2) and Engineering (1).

**Table 1**: Twenty-three LIS authors listed among the most highly-cited authors at http://highlycited.com/ in the ESI broad categories "Social Science, general," "Economics & Business," "Computer Science" and "Engineering."

| First name | Surname | ESI broad category |
|---|---|---|
| Nancy E | Adler | Social Science, general |
| Neeraj K | Arora | Social Science, general |
| Joan S | Ash | Social Science, general |
| David W | Bates | Social Science, general |
| Lutz | Bornmann | Social Science, general |
| Hans-Dieter | Daniel | Social Science, general |
| Leo | Egghe | Social Science, general |
| Tejal K | Gandhi | Social Science, general |
| Wolfgang | Glänzel | Social Science, general |
| Russell E | Glasgow | Social Science, general |
| Ashish K | Jha | Social Science, general |
| Gilad J | Kuperman | Social Science, general |
| Loet | Leydesdorff | Social Science, general |
| Seth M | Noar | Social Science, general |
| Ismael | Rafols | Social Science, general |
| Ronald | Rousseau | Social Science, general |
| Anthony F J | Van Raan | Social Science, general |
| Richard B | Bagozzi | Economics & Business |
| Paul A | Pavlou | Economics & Business |
| Visvanath | Venkatesh | Economics & Business |
| Isaac S | Kohane | Computer Science |
| Marylyn D | Ritchie | Computer Science |
| Enrique | Herrera-Viedma | Engineering |

Furthermore, we counted the number of highly-cited papers for each of the 798 authors (Figure 1). Twenty-one of the 798 LIS authors contributed to four or more papers (see Figure 1 for the names of these authors). Comparing Table 1 and Figure 1 reveals that a high percentage (57%) of the authors, who contributed four or more papers to the 305 top-1% highly-cited LIS papers, are also listed at http://highlycited.com/. A similar overlap can be observed, when another percentile rank is chosen: using the top-1‰ most frequently cited papers of the WoS set, for example, 71 authors are listed, among whom five co-authored more than a single paper in the set. But nine of the 23 authors (39%) listed in Table 1 are on the top-1‰ list and four of them contribute more than a single paper to the underlying 28 papers. In other words, most highly-cited authors appear to emerge on different lists even if other parameters are applied to the analysis.

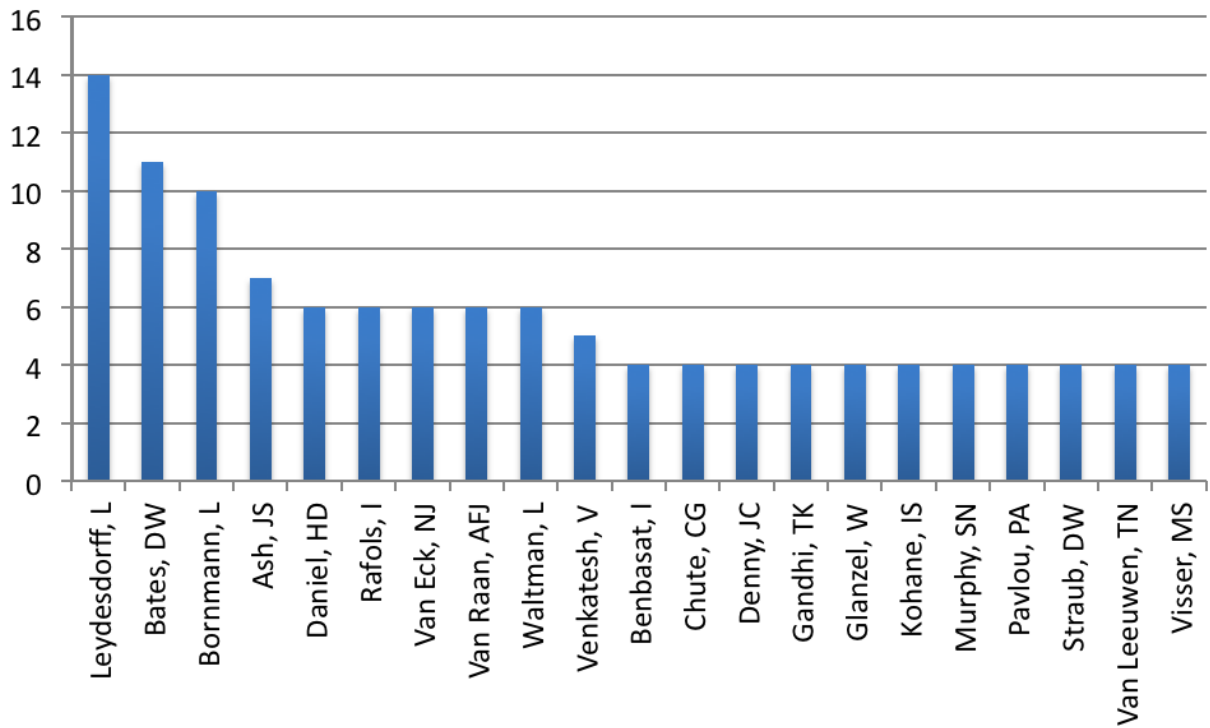

**Figure 1**: Twenty-one authors who contributed four or more top-1% highly-cited papers in the LIS category.

*3.2 Collaboration of authors with highly-cited papers*

Figure 2 shows eight groups with ten or more co-authors among the 798 co-authors of the top-1% highly-cited papers in the LIS subset. The eight groups together comprise 347 authors, who contributed to 113 of the 305 papers under study. Within this domain of 305 highly-cited papers, these eight groups are no longer connected among them in terms of co-authorship relations.

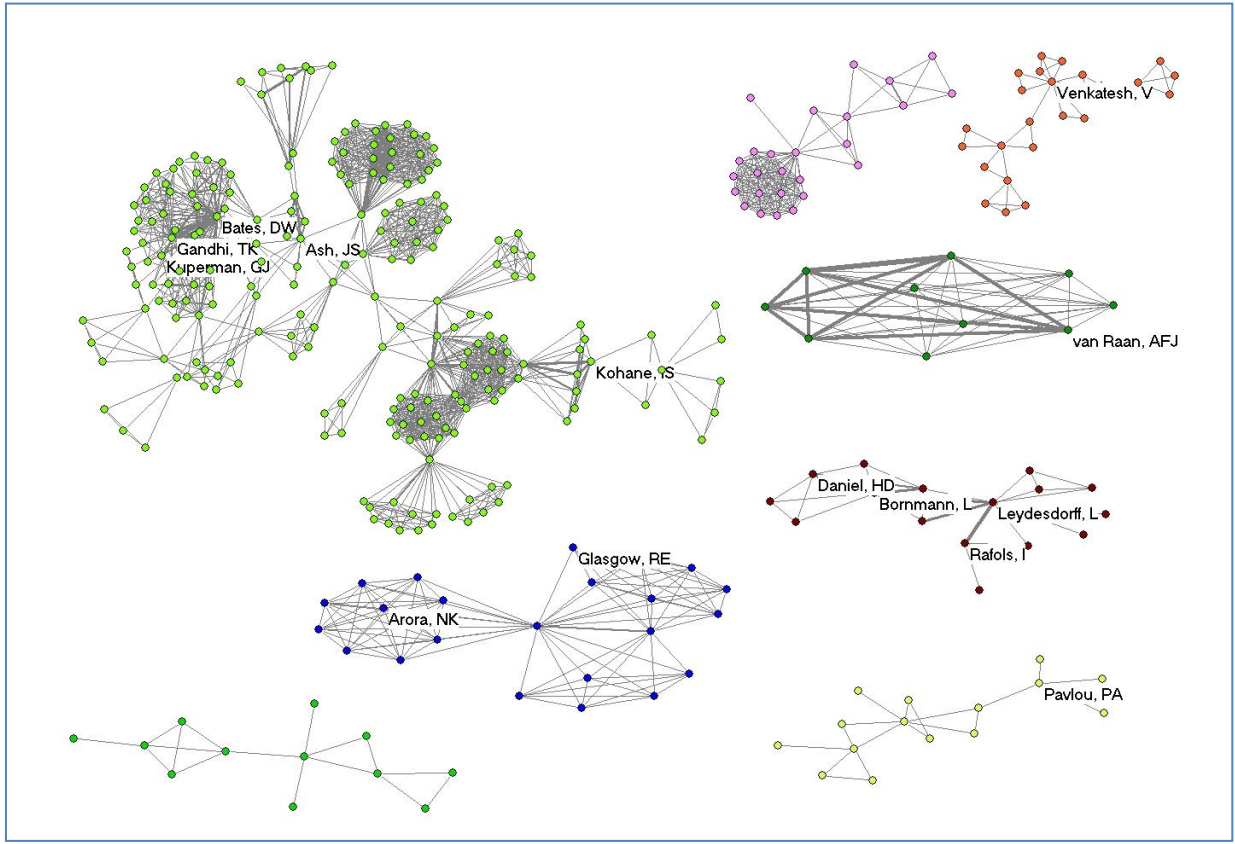

**Figure 2**: Networks of collaborating authors. Author names are indicated for authors included in Table 1.

Of the eight groups shown in Figure 2, the largest component contains 220 co-authors (27.6%) including the most highly-cited authors Bates, Ghandi, Kuperman, Ash, and Kohane (see Table 1). The affiliations of the majority of these authors are located in the USA. In Europe, one group of authors is affiliated to the Erasmus University of Rotterdam. Together these authors in this component published 45 top-1% highly-cited papers. All these publications appeared in journals assigned not only to the WoS category "Information Science & Library Science," but also to the WoS category "Medical Informatics." About half of the papers report on clinical trials and drug related information as well as decision support systems for physicians. In other papers, the possibilities for the smart use of computers and electronic information in clinical practice are reported.

A second group of authors also study medicine-related topics. This group comprises 24 (co)-authors, whose affiliations are mainly located at the NIH and a few other American institutions. Among them are the most-highly cited authors Glasgow and Aurora. Together the 24 authors co-authored three highly-cited publications. These three publications appeared in journals assigned to LIS and also to the categories "Communication" (2) or "Medical Informatics" (1). The studies focus on the collection of patient data and their exploitation.

A third group of scientists (indicated by the pink spheres at the upper edge of Figure 2) wrote highly-cited papers about medical topics. The group consists of 26 (co)-authors, whose affiliations are distributed over the USA. Of the six papers written, five appeared in journals assigned to both LIS and the category "Medical Informatics." All the papers deal with the exploitation of recorded medical and biochemical data in health care systems.

Three other circles of collaborating authors of highly-cited papers investigated several aspects of the internet. One group includes the most highly-cited author Venkatesh; it consists of 25 (co)-authors. The affiliations of the majority of these authors are located in the USA, but two authors use affiliations in Hong Kong and one in Singapore. They co-authored 12 highly-cited publications which appeared in journals assigned to the category LIS and the category "Management." In these papers, the focus is on how people react when they are confronted with modern information technology in management and business.

Another group consisting of 15 (co)-authors includes the most highly-cited author Pavlou. Their affiliations are distributed over the USA and France. These authors wrote ten highly-cited publications, which appeared in journals assigned to LIS and "Management" as categories. The majority of these papers deal with the application of information technologies in business and commerce.

A sixth circle of 11 (co)-authors (indicated by green spheres at the lower edge of Figure 2) published seven highly-cited publications in journals, which are all assigned to the categories LIS and "Management." In four of these papers Benbasat was involved as a co-author. Most of the affiliations of these authors are to addresses in Canada, while a minority of the authors works in Asia. The common interest of the co-authors is based on computers, internet, and issues of social psychology.

Two smaller groups are recognizable as specialized in scientometrics: one around Van Raan as the senior author of the Centre for Science and Technology Studies (CWTS) of Leiden University, and one European network of several senior authors (Bornmann, Daniel, Leydesdorff, and Rafols). Various aspects of scientometrics such as performance indicators or citation-based rankings are subjects of their highly-cited papers.

The detailed evaluation of Figure 2 suggests that it is informative to consider the second or third categories (in addition to LIS), to which the journals of the highly-cited papers are assigned. As summarized in Table 2, it then becomes apparent that the highly-cited papers focus on topics such as: i) recording and exploiting of medical information in health care and clinical praxis; ii) success of the application of information technology in commerce and business; iii) performance measurement of scientific publications. Common interests in these topics appear to be a driving force of collaboration among highly-cited authors.

**Table 2:** All the journals, in which the 113 highly-cited papers of the eight groups of authors shown in Figure 2 have been published, were assigned to the WoS category "Information Science & Library Science." Many of these journals were assigned to additional categories as indicated below. The numbers in brackets indicate the number of papers in the respective categories.

| No. of highly-cited papers in LIS | Most frequent additional WoS category | | Second most frequent additional WoS category | |
|---|---|---|---|---|
| 45 | Medical Informatics | (45) | Comp.Sci.Interdisc.Appl. | (45) |
| 3 | Communication | (2) | Medical Informatics | (1) |
| 6 | Comp.Sci.Interdisc.Appl. | (5) | Medical Informatics | (5) |
| 12 | Management | (12) | Comp.Sci.Inform.Systems | (10) |
| 10 | Management | (10) | Comp.Sci.Inform.Systems | (6) |
| 7 | Comp.Sci.Inform.Systems | (7) | Management | (6) |
| 22 | Comp.Sci.Inform.Systems | (14) | Comp.Sci.Interdisc.Appl. | (4) |
| 8 | Comp.Sci.Interdisc.Appl. | (4) | Comp.Sci.Inform.Systems | (2) |

*3.3 Institutional contributions to the 305 most highly-cited papers in LIS*

In many highly-cited papers not just a single, but several institutions from different authors are listed as institutional affiliations. Furthermore, several authors do not mention only a single, but up to three affiliations. For this reason, we produced four ranking lists, which include the institutional affiliations of authors in different ways.

The first ranking list of institutions (see Table 3A) is based on the authors' primary affiliations. All primary affiliations of authors are whole-number counted: Each mention of a unique institution on a paper leads to a full point. The second ranking is also based on the primary affiliations of authors (see Table 3B). However, the institutional affiliations are fractionally counted. Using the method of fractional counting, the number of unique affiliations provided on a paper is taken into account: If three primary institutions are mentioned by an author, for instance, each institution is counted as 1/3.

As the results in Table 3 show, whole-number and fractional counting lead to different results. While whole-number counting shows that Harvard University (n=18), Brigham and Women's Hospital and the University of Amsterdam (each n=15) have the largest numbers of most highly-cited papers, Leiden University (n=10.50) followed by the University of Amsterdam (n=9.42) are leading the rank established using fractional counting. Depending on the kind of counting 16 or 17 institutions in the USA are among the top-20 institutions.

**Table 3:** Numbers of highly-cited papers per institution, determined by the authors' primary institution using either the whole-number counting (A) or the fractional counting method (B). The ten institutions with the highest number of highly-cited papers are shown, respectively.

| Rank | Institution | Numbers of highly-cited LIS papers |
|---|---|---|
| A) whole-number counting: | | |
| 1. | Harvard University | 18 |
| 2. | Brigham and Women's Hospital | 15 |
| 3. | University of Amsterdam | 15 |
| 4. | University of California System | 14 |
| 5. | Partners Healthcare System | 12 |
| 6. | Leiden University | 11 |
| 7. | Vanderbilt University | 11 |
| 8. | Georgia State University | 10 |
| 9. | Indiana University | 10 |
| 10. | Oregon Health and Science University | 10 |

B) fractional counting
| | | |
|---|---|---|
| 1. | Leiden University | 10.50 |
| 2. | University of Amsterdam | 9.42 |
| 3. | University of California System | 9.03 |
| 4. | Harvard University | 7.95 |
| 5. | Vanderbilt University | 5.99 |
| 6. | ETH Zürich | 5.83 |
| 7. | Indiana University | 5.20 |
| 8. | Columbia University New York | 4.93 |
| 9. | University of Maryland | 4.92 |
| 10. | Brigham and Women's Hospital | 4.91 |

______________________________________________________________________

In addition to the lists shown in Table 3, we calculated two other rankings based on all the institutions named by the authors (primary and further addresses). However, the consideration of these further addresses did not lead to significantly changed results. When whole-number counting of primary and further addresses was applied, University of Amsterdam and University of California System as well as Leiden University and Vanderbilt University changed their relative positions. Similar minor changes of ranks were observed, if fractional counting was applied additionally. Using the LIS dataset in this study, we did not find an unexpected and rather unknown university as Bornmann & Bauer (2014) described for the case of the King Abdulaziz University in Saudi Arabia based on the dataset at http://highlycited.com.

*3.4 Collaboration between different institutions*

Generation of institutional networks revealed two groups four or more institutions (figure not shown). Together, they contain 186 of the 275 institutions (67.6%). One giant component includes 181 institutions of mainly American universities, which are highly interconnected. Canadian universities are linked into the main component via the University of British Columbia.

Their network is based on studies about the use and acceptance of the internet. European universities are especially visible for authors interested in scientometrics. One of them, the University of Amsterdam, is connected to American partners and to the ETH Zürich. But the Leiden University links exclusively to European centers forming a second independent group of five institutions.

## 4. Discussion

This study deals with highly-cited papers, their authors and affiliations in LIS. Being (co)-author of one highly-cited paper rarely qualifies for an inclusion in a top-ranked list of authors. In order to find the authors of the 305 highly-cited papers, who have particular impact on research in the field of LIS, we determined those authors who contributed four or more papers to the 305 highly-cited LIS papers or more (Figure 1) than a single paper to the 28 top-1‰ LIS papers. In addition, we compared the names with those who were listed as most highly-cited author at http://highlycited.com after a more rigorous selection process (Table 1).

Twenty-three of the 798 LIS-authors were also found on the highlycited.com list. Twelve of them contributed four or more papers to the papers analyzed in this study. Nine are also authors of the top-1‰ papers among the LIS publications. Nine further authors do not belong to the most highly-cited authors at highlycited.com, but contributed four or more papers to the 305 highly-cited LIS papers. Another author contributed with more than a single paper to the top-1‰ LIS papers. In summary, 33 authors complied with at least one selection criterion of this study. Twelve of them complied with two criteria: but four fulfilled all three criteria. Interestingly, all the 33 authors complying with at least one of our criteria belong to one of the networks of collaboration shown in Figure 2. As collaboration often points to innovative work (Lee & Bozeman, 2005; Bidault & Hildebrand, 2014), we evaluated the papers published by the eight collaboration circles (Figure 2) in detail.

Looking at the institutions, to which the various authors are affiliated, it became obvious that most of them are located in the USA. Among the 20 top universities no more than four universities are located in Europe, independently whether whole-number or fractional counting is applied. Even if the second or third affiliation provided by the authors were considered, the ranking of the institutions changed only marginally. There is one exclusive circle of collaboration of five institutions in Europe. Its center is Leiden University, which is leading the ranking in the case of fractional counting.

The LIS category is a mixed bag of journals in bibliometrics, library science, management and medical information systems with different citation characteristics (Leydesdorff & Bornmann, 2011; Milojević & Leydesdorff, 2013). LIS journals are used as publication outlets by scientists working in the social sciences and on business topics; but information specialists in computer science and technology also contribute. The share of the various topics in the total number of manuscripts varies from year to year (Tang, 2014).

Between 2002 and 2012, the shares of papers of journals belonging to LIS and in addition to medical or management categories in total number of LIS articles and reviews increased from 3,6 to 4,8 % and from 6,2 to 10,1%, respectively. However, the number of top-1% papers in journals belonging to the LIS and Medical Information categories increased from two in 2002 almost linearly to 15 in 2012, while at the same time the number of highly-cited papers in management journals decreased from 21 to five per year. Focusing on papers published by circles of collaborating top-1% authors (Figure 2, Table 2) makes the change of interest in these topics even more pronounced, as 51 top-1% papers were found in medical related and 28 in management related journals. Hence we conclude that more information about trends in research could be gathered by identifying top-1% papers and collaboration circles publishing top-1% papers than when counting publications. Future work will show whether this way of analyzing top-1% papers can indicate new trends in other research areas as well.

**Acknowledgements**

The bibliometric data used in this paper are from an in-house database developed and maintained by the Max Planck Digital Library (MPDL, Munich) and derived from the Science Citation Index Expanded (SCI-E), Social Sciences Citation Index (SSCI), Arts and Humanities Citation Index (AHCI) prepared by Thomson Reuters (Scientific).

**Footnotes**

[1] The same search at the WoS interface provided 31,443 records on this date. However, Thomson Reuters regularly updates the database with previous volumes of newly added journals, whereas MPDL was dated at the end of 2013.

[2] Pajek is a program for network analysis and visualization, which is freely available for academic purposes at http://pajek.imfm.si/doku.php?id=download (de Nooy, Mrvar, & Batagelj, 2011).